# Characterization of Starlink Direct-to-Cell Satellites In Brightness Mitigation Mode

Anthony Mallama*[1], Richard E. Cole[1], Jay Respler[1] and Scott Harrington

2025 January 14

[1] IAU - Centre for the Protection of Dark and Quiet Skies from Satellite Constellation Interference

* Correspondence: anthony.mallama@gmail.com

Abstract

The mean apparent magnitude of Starlink Mini Direct-To-Cell (DTC) satellites observed in brightness mitigation mode is 5.16, while the mean of magnitudes adjusted to a uniform distance of 1,000 km is 6.47. The DTCs have faded since early in 2024 because SpaceX subsequently adjusted the spacecraft attitudes to dim them. A physical model for satellite brightness that fits the observations is described.

1. Introduction

Bright satellites interfere with observation of the night sky. So, they are a problem for research astronomers (Barentine et al. 2023) and a distraction for casual observers (Mallama and Young 2021).

SpaceX began launching Starlink satellites equipped for Direct-To-Cell (DTC) communication in January 2024. These spacecraft orbit below the internet satellites which renders them more luminous, and they have a different form factor that includes the DTC antenna.

We previously analyzed the brightness of these DTC spacecraft using observations recorded through June 2024 (Mallama et al., 2024). Their mean apparent magnitude was 4.62 while the mean of magnitudes adjusted to a uniform distance of 1,000 km was 5.50. At that time DTCs averaged 4.9 times brighter than Starlink Mini internet spacecraft at a common distance.

Meanwhile, SpaceX informed us that the DTC satellites were still undergoing testing and had not yet been placed in brightness mitigation attitudes when we obtained our data. That statement motivated this study which reports on the brightness of DTCs from July through December 2024.

Section 2 describes how magnitudes were determined for this research. Section 3 characterizes the brightness of DTC satellites. Section 4 describes a physical model of spacecraft luminosity. Section 5 discusses the DTC impact on astronomy and Section 6 summarizes our findings.

2. Observational methods

Magnitudes were recorded using electronic and visual techniques. Electronic measurements were obtained at the MMT9 robotic observatory (Karpov et al. 2015 and Beskin et al. 2017). The MMT9 hardware consists of nine 71 mm diameter f/1.2 lenses and 2160 x 2560 sCMOS sensors. MMT9 photometry is within 0.1 magnitude of the V-band according to information in a private communication from S. Karpov as discussed by Mallama (2021).

We obtained apparent magnitudes from the MMT9 on-line database along with ranges (distances between the spacecraft and the observer) and phase angles (the arcs measured at the satellite between directions to the Sun and the observer). MMT9 logs data at 10 Hertz cadence and we averaged those into 5 second mean magnitudes for this analysis.

For visual observations, spacecraft brightness is determined by comparison to nearby reference stars. The resulting magnitudes approximate the V-band. Angular proximity between satellites and stellar objects accounts for variations in sky transparency and sky brightness. This method of observing is described in detail by Mallama (2022).

A total of 551 DTC observations from MMT9 and visual observers are used in this study. They are available from SCORE, the CPS database.

3. Brightness characterization
The distribution of apparent magnitudes for DTC satellites is shown in Figure 1. Early (before 2024 June 30) and late observations are distinguished and they are compared to magnitudes for Starlink internet satellites. The distribution of late DTC magnitudes is distinctly fainter than that of early ones.

The apparent brightness of late observations peaks at magnitude 5.5. The peak of early observations was the same but that distribution is skewed much more strongly toward brighter values. The peak for internet satellites is 6.5.

The mean apparent magnitude for late DTCs is 5.16 with a standard deviation (SD) of 1.30 and a standard deviation of the mean (SDM) of 0.06. The corresponding values for internet Minis are 6.36, 0.63 and 0.01. DTCs are more luminous and their brightness dispersion is larger.

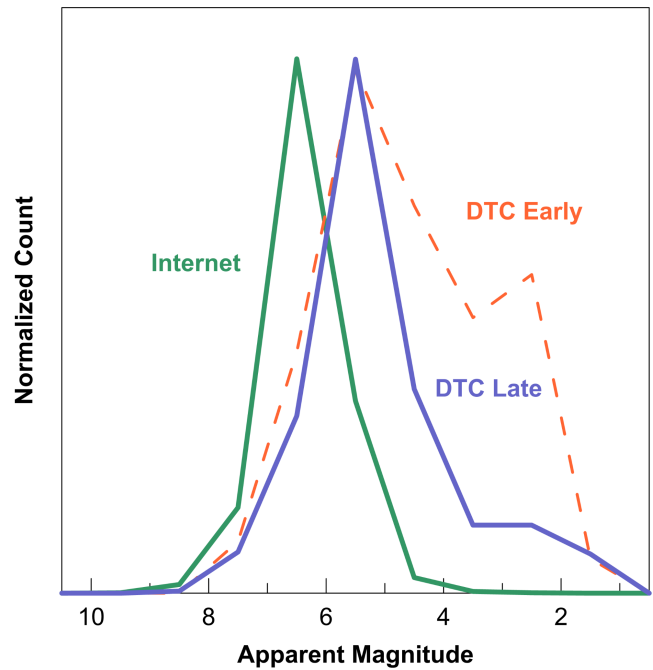

Figure 1. The distribution of apparent magnitudes for Starlink Mini DTC (early and late observations) and internet satellites.

Adjusting apparent magnitudes to a common distance is useful for comparing different populations of satellites. This applies to Starlink Minis because the internet satellites orbit at about 550 km while the DTCs are around 200 km lower. (At the time of this writing SpaceX has begun placing internet satellites into 450 km orbits. Those spacecraft are not considered in this study.)

When apparent magnitudes for late DTC observations are adjusted to a uniform distance of 1,000 km their mean, SD and SDM are 6.47, 1.32 and 0.06. The corresponding values for internet spacecraft are 7.22, 0.83 and 0.01. The difference between the mean magnitudes indicates that DTCs are 2.0 times brighter than the internet spacecraft when observed at the same distance. The distributions are shown in Figure 2.

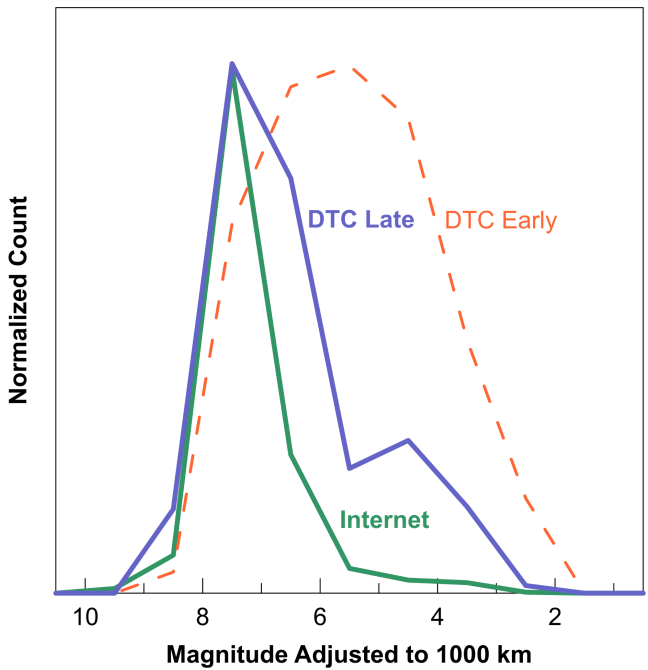

*Figure 2. The distribution of magnitudes adjusted to a distance of 1,000 km for Starlink Mini DTC (early and late observations) and internet satellites.*

Satellite brightness is affected by the phase angle which is the arc measured at the satellite between directions to the Sun and to the observer. Phase angle is the independent variable of the phase function where 1000-km magnitude is dependent.

The phase functions for Starlink DTC and internet satellites display a concave upwards shape as shown in Figure 3. The high luminosity at small phase angles occurs because the satellites are nearly opposite the Sun from the observer and are almost fully lit. Meanwhile, the brightness at large phase angles occurs when the spacecraft are between the Sun and the observer. Then the high luminosity indicates forward scattering from components that are mostly back-lit and also additional brightness contributions discussed later.

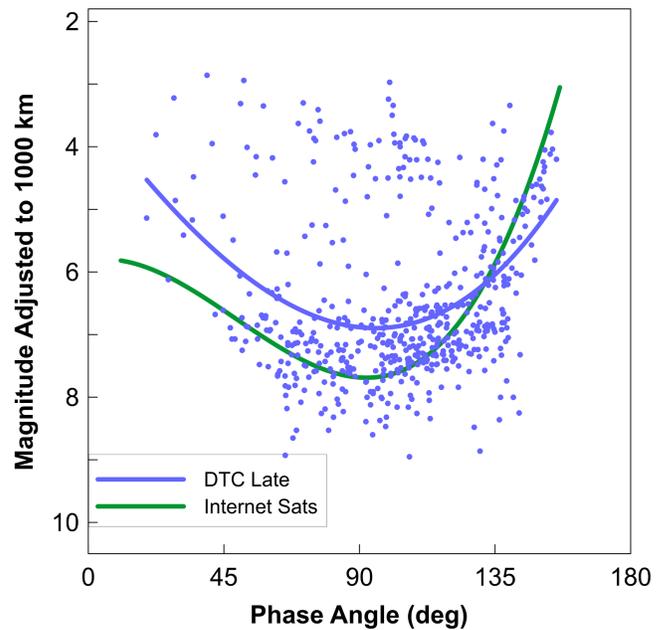

*Figure 3. The phase functions for late DTC observations and for internet satellites.*

Table 1. Phase Function Polynomial Coefficients

```
    Order->     0        1         2         3
DTC Late      3.365   0.06376  -1.716E-4  -1.139E-6
Internet      5.822  -0.00879   8.483E-4  -5.784E-6
```

4. Physical model

4.1 Appearance on-orbit

Figure 4 shows a stack of Starlink spacecraft with a Starlink DTC on top of the stack. The large DTC antenna is seen folded and indicates the antenna surface is reflective, but not as perfectly reflective as the chassis base of the spacecraft as has been seen in other images of Starlink stacks. The DTC antenna in the image appears to have several sections that will unfold after launch.

Figure 5 is a telescopic image of a DTC spacecraft on-orbit taken by Tom Williams from the UK. The bottom pane gives our interpretation of the image.

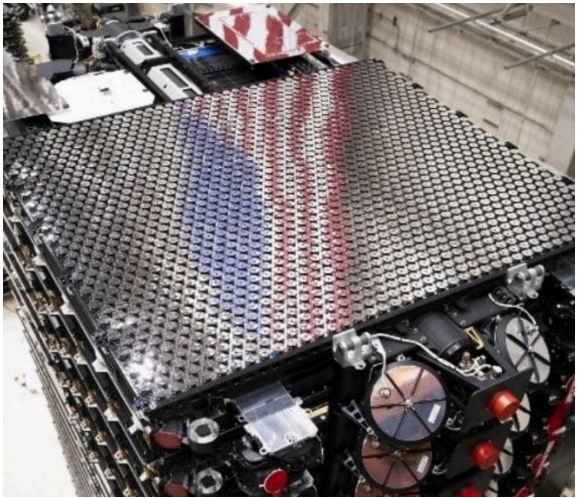

*Figure 4. A stack of folded Starlink spacecraft before launch, a DTC Starlink on top (Image from SpaceX).*

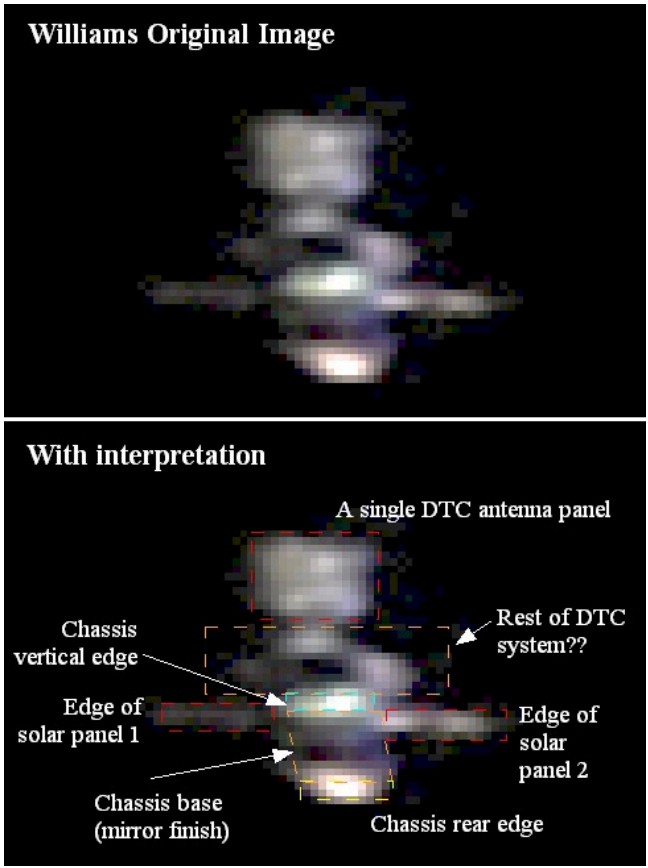

*Figure 5. Telescopic image of a Starlink DTC spacecraft with interpretation (Image courtesy of Tom Williams, Wiltshire, UK).*

The mirror-surfaced chassis of the spacecraft is dark as it is reflecting the dark night-time Earth's surface. The front and rear edges of the chassis are not mirror-surfaced (though they are black) and reflecting the Sun. The solar panels are rotated so only the edge is catching the Sun. Space-X have published that for brightness mitigation reasons this is the normal orientation of the Gen2 Starlink solar panels when the spacecraft are crossing the terminator and thus visible from the ground. They added that this panel orientation is not always possible for operational reasons so the panels may be bright at some times. The unfolded DTC antenna is visible at the top of the image and is expected to be maintained parallel to the surface of the Earth, as is the chassis of the spacecraft.

Thus, the optical signature of the spacecraft can be represented (approximately) as a flat earth-facing surface (the DTC antenna and the chassis base) and two vertical surfaces (the front and rear vertical edges of the chassis). The reflection from the solar panels is considered to be suppressed by the operational procedures.

Space-X stated that the DTC spacecraft can be aligned with the orbit velocity-vector in a number of ways, depending on operational requirements. We have assumed here that the long axis of the spacecraft, through the DTC antenna, remains oriented parallel to the orbit velocity vector.

4.2 The optical model

Based on previous work on Gen1 Starlinks (Cole 2021), a simple numerical model was constructed on the basis described above. All reflections are considered to be Lambertian or diffuse reflection and allowance is made for the orientations of the three surfaces with respect to the Sun and to the observer. The solar panel is not separately included in the model.

The differences in magnitude between the model and the July to December 2024 observations are calculated. The overall predicted brightness of the spacecraft is normalized to match the overall observed brightness. The average relative contribution of the vertical and horizontal surfaces is also fitted to the data, so there are a total of two fitted parameters in the model.

The residuals between the first, simple model and the observations are shown in Figure 6, plotted against phase angle. The visual observations and MMT9 observations are plotted separately. Note that MMT9 does not observe at high phase angle, close to the Sun. This plot reveals two things:

1. There are a set of observations that are much brighter than the majority. The observers often noted the spacecraft had a blue color at these times, hence the term Blue Zone in Figure 6. It is thought that the solar panel is not in the low-brightness orientation and that the panel has a blue color (as can also be seen in the few available images from on-board Starlink cameras).
2. The simple model greatly under-predicts the observed magnitude at phase angles greater than 110°. This is due to direct specular reflection of the bright, sunlit surface of the Earth by the mirror surface of the chassis and, less perfectly, by the reflective surface of the DTC antenna.

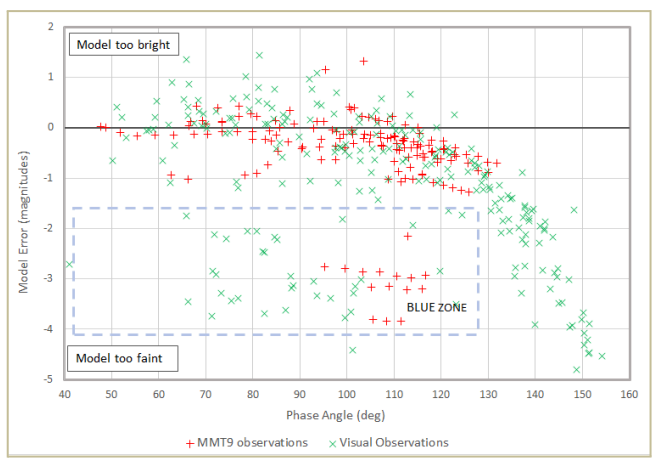

Figure 6. Errors from the simple model, plotted against phase angle. MMT9 and visual observations are plotted separately.

The bright earth reflection is also observed in the non-DTC Starlinks. The Earth-facing mirror surfaces are reflecting the Earth's surface thousands of kilometers from the observer, where the Sun is above the horizon. It can be well-modeled as a function of the sun elevation at that point on the Earth. This contribution is significant, as can be seen in Figure 3 and is the cause of the upward turn in the phase function at high phase angle.

There will be a small, additional contribution from reflection of earthlight by the non-mirrored surfaces on the spacecraft. Earthlight is diffuse reflected sunlight, mostly from the Earth's surface directly below the spacecraft. In the case of Starlinks, with their mirror surfaces, this is much smaller than the directly reflected bright earth reflection.

The residuals from the model with an additional bright earth term are shown in Figure 7. The fit is not perfect, but the brightness of the sunlit Earth's surface is not uniform given different weather conditions and surface structure (e.g. land, sea, ice). The area of the Earth being reflected, at any instant, by the mirror surfaces is very small, tens of meters across, so its brightness contribution will vary.

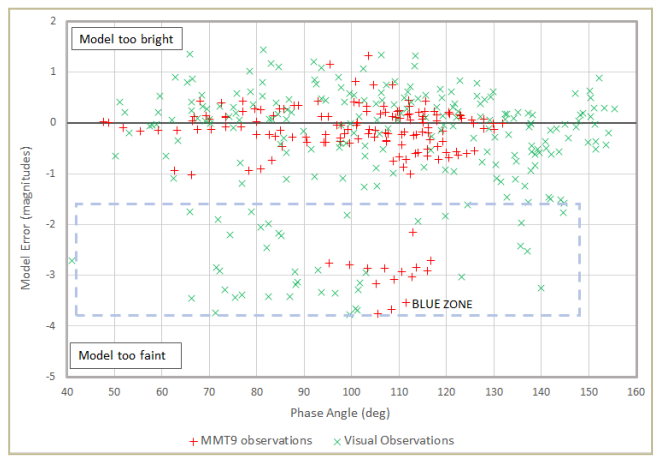

Figure 7. Errors from the model with bright Earth contribution, plotted against phase angle

Finally, Figure 8 shows the residuals from the model against the Sun elevation at the spacecraft, that is the angle of the Sun on the underside of the chassis. Given that SpaceX have said the orientation of the DTC spacecraft with respect to the velocity-vector is variable (and not published), the trend of the errors in Figure 8 is reasonably close to the x-axis. The model can be used to generate predictions of the DTC spacecraft brightness across the sky for any sun elevation and azimuth at the observer.

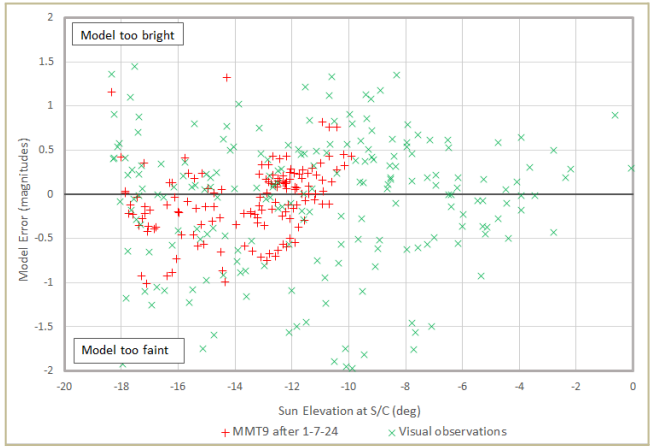

*Figure 8. Errors from the model with bright Earth contribution, plotted against elevation of the Sun at the spacecraft*

5. Discussion
In this section we briefly summarize the effect of DTC satellites on astronomy. Then we offer a hypothesis to explain the skew of their magnitude distribution.

Satellites brighter than magnitude 7 impact research astronomy (IAU, 2024) by seriously contaminating photographic images. For casual sky watchers spacecraft brighter than magnitude 6 are a distraction because they are visible to the unaided eye. DTC spacecraft are brighter than these limits, so they will adversely impact astronomical research and aesthetic appreciation of the night sky.

The DTC magnitude distributions are skewed toward brighter values as shown in Figures 1 and 2. For apparent magnitudes the excess begins at magnitude 3.5, and for 1000-km values it starts at magnitude 6.0. Meanwhile, the distribution for internet satellites are more nearly symmetrical.

The explanation for the skew of DTC brightness distribution may be that these spacecraft are removed from brightness mitigation mode more often than internet satellites in response to the greater atmospheric drag at their lower altitudes.

Atmospheric density is a function of height and it also depends on the level of solar activity. The ratio of densities at the heights of DTCs (near 350 km) and internet satellites (near 550 km) was derived from the MSISE-90 model. The ratio is about 30 for mean solar activity while it is about 10 for extremely high activity. The observations in this report were obtained when solar activity was moderately high, so we take the ratio to be 20.

Drag is proportional to density times velocity squared, and the velocity ratio for 340 and 550 km is near unity (viz., 1.015). So, the drag ratio is still about 20.

High drag requires a greater frequency of station-keeping maneuvers. Starlink satellites at all altitudes are temporarily removed from their brightness mitigation attitude during maneuvers. So, those at lower altitudes would brighten more often due to the more frequent suspension of mitigation during maneuvering.

Evidence for more frequent maneuvering is shown in Figure 9. The top panel plots revolutions per day versus day of year for internet Starlink number 30509 at 560 km, while the bottom panel shows the same for DTC Starlink 11072 at 360 km. The vertical scales are equal. The RMS residual for the DTC is more than twice that of the internet satellite. The greater variation for the DTC spacecraft suggests that station keeping occurs more frequently.

The frequent maneuvers could also make the satellites more difficult for astronomers to avoid in their observing plans. Published orbital elements will represent the orbit less accurately because they become outdated more often.

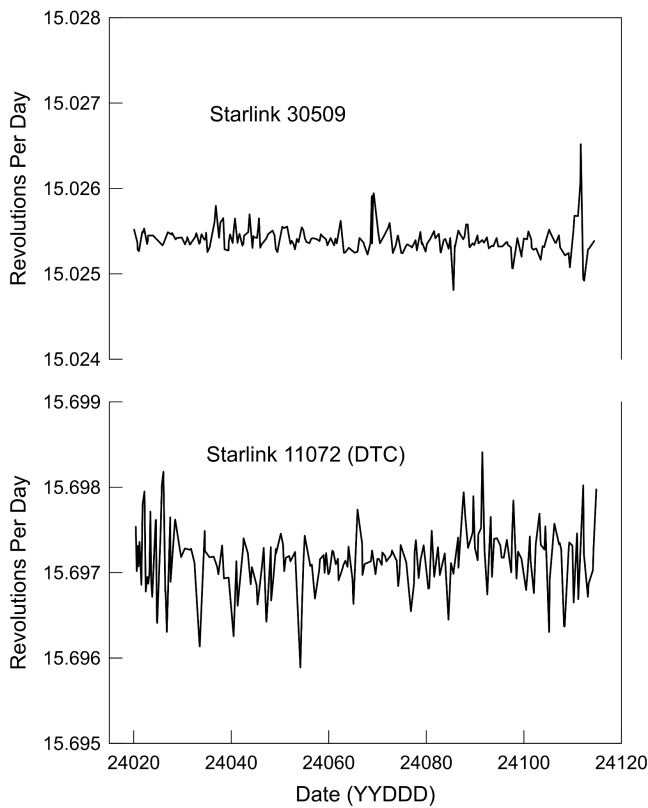

*Figure 9. Variations in the revolutions per day for an internet satellite (top) and a DTC spacecraft (bottom).*

6. Conclusions
The mean apparent magnitude of Starlink DTC satellites is 5.16 while the mean of magnitudes adjusted to a uniform distance of 1,000 km is 6.47. They average 2.0 times brighter than Starlink internet spacecraft when observed at a common distance. A physical model for satellite brightness fits the observations. DTC satellites negatively impact professional astronomy and aesthetic appreciation of the night sky.

Acknowledgements
SpaceX provided important information about the status of brightness mitigation for DTC satellites. We thank the staff of the MMT9 robotic observatory for making their data available. We also thank Tom Williams for his on-orbit image of a DTC Starlink. The Heavens-Above.com web-site was used to plan observations. Stellarium, Orbitron, Heavensat and Cartes du Ciel were employed to process the resulting data.